\documentclass[aps,prb,twocolumn,showpacs,floatfix,superscriptaddress]{revtex4}
\usepackage{amsmath}
\usepackage{amsfonts}
\usepackage{amssymb}
\usepackage{euscript}
\usepackage{bm}
\usepackage{epsfig}
\usepackage{graphicx}
\usepackage{color}

\begin{document}

\title{Quasiparticle properties of an impurity in a Fermi gas}
\author{Jonas Vlietinck}
\author{Jan Ryckebusch}
\author{Kris Van Houcke}
\affiliation{Department of Physics and Astronomy, Ghent University, Proeftuinstraat 86, 9000 Gent, Belgium}

\begin{abstract}
We report on a study of a spin-down impurity strongly coupled to a
spin-up Fermi sea (a so-called Fermi polaron) with the diagrammatic
Monte-Carlo (DiagMC) technique. Conditions of zero temperature and
three dimensions are considered
for an ultracold atomic gas with resonant interactions in the zero-range limit.
 A Feynman diagrammatic series is developed
for the one-body and two-body propagators providing information about
the polaron and molecule channel respectively. 
The DiagMC technique allows us to reach diagram orders that are high enough for extrapolation to infinite order.
The robustness
of the extracted results is examined by checking various resummation
techniques and by running the simulations with various choices for the
propagators and vertex functions. 
It turns out that dressing the lines in the diagrams as much as possible is not always the optimal choice.
We also identify classes of dominant diagrams for the one-body and two-body self-energy in the region of strong interaction. These dominant diagrams turn out to be 
 the leading processes of the strong-coupling limit. 
The quasiparticle energies and
$Z$-factor are obtained as a function of the interaction strength. We
find that the DiagMC results for the molecule and polaron properties 
are very similar to those obtained with a variational ansatz.
 Surprisingly, this variational ansatz gives very good predictions for the quasiparticle residue even when this residue is significantly smaller than one.
\end{abstract}


\maketitle

\section{Introduction}

The notion of a `bare' particle loses its significance once it is strongly coupled to a medium. 
Landau introduced the notion of 
a quasiparticle whose properties may be very different from those of a bare particle \cite{landau}. 
The most prominent example is an electron moving in a 
crystal: the electron displaces the nearby ions and carries this 
distortion with it.  The presence of the phonon cloud  changes  the mass and energy of the electron, that is  dubbed as `polaron'  \cite{LLStatMech2}. 
More generally, a polaron arises whenever a quantum impurity is strongly coupled 
to an environment. 
 These quantum-mechanical quasiparticles play a key role in the low-energy behavior 
 of a macroscopic quantum liquid. 

In recent years,    the field of ultracold atoms has  provided an exciting framework 
for studying polaronic effects. 
A key idea is that models designed for describing the rich and non-trivial structure of the solid state, can 
be emulated in a clean and controllable manner with ultracold atoms. 
For example,  so-called Fermi polarons \cite{polaron1,schirotzek,nascim}, spin-down impurities that are strongly coupled to a spin-up Fermi sea, can be created in
 a degenerate two-component  atomic Fermi gas 
  when going to the limit of strong spin-imbalance  close to a Feshbach resonance. 
  The impurity is  coherently dressed with particle-hole excitations of the Fermi sea. 
  The
  properties of the Fermi polaron are important for the quantitative understanding 
 of a strongly imbalanced Fermi gas \cite{pilati}.
 
 In this paper, we focus on the  `attractive Fermi polaron',  with an  attractive interaction between the impurity and the fermions of the bath. 
A recent experiment using  an ultracold gas of $^6$Li atoms in three dimensions revealed  the existence of Fermi polarons through a narrow quasiparticle peak in the impurities'  radio-frequency (rf) spectrum \cite{schirotzek}.  
At a critical interaction strength, the  disappearance of this peak was interpreted as a transition from polaronic to molecular binding, when the impurity and an atom of the sea form a two-body bound state. 
Such a transition had
theoretically been predicted in three dimensions (3D) by   Prokof'ev and Svistunov \cite{polaron1}. 
To determine the transition point, 
they developed a diagrammatic Monte Carlo technique (DiagMC) capable of solving the Fermi polaron model\cite{polaron1,polaron2}.
Calculations of the ground-state energy showed that for a sufficiently strong attraction between the impurity atom and the atoms of the spin-up Fermi sea, a molecular state becomes energetically favorable.  The crossing point was found at an interaction strength $(k_F a)_c = 1.11(2)$, with $k_F$ the Fermi momentum of the spin-up sea and $a$ the $s$-wave scattering length. A  variational treatment developed by Chevy based on an expansion up to single particle-hole excitations on top of the unperturbed Fermi sea turned out to be remarkably accurate \cite{chevy}. 
A combination of Chevy's ansatz with a variational wave function in the molecular limit \cite{mora,punk,mol3} also revealed the polaron-to-molecule transition,  very close to the DiagMC result.

In the present work, 
we study the quasiparticle properties of the Fermi polaron problem in 3D with the DiagMC technique \cite{polaron1,polaron2}.  
This technique
evaluates a series of Feynman diagrams for the one-particle and two-particle proper self-energies.  
A full description of the DiagMC algorithm was presented  in Ref.~\cite{polaron2}. 
Building on the work of Ref.~\cite{polaron2} we have implemented the 
DiagMC algorithm independently.
We explore various DiagMC schemes \cite{bold} and  series resummation methods to check the robustness of the results against the possible uncertainties of summing the series. 
First, we  confirm the transition point.  Next, we calculate the quasiparticle residue which we compare to experimental data and variational results. The quasiparticle residue, or $Z$-factor, gives the overlap of the non-interacting wave function and the fully interacting one,
\begin{equation}
Z_p =    | \langle \Psi^{N_{\uparrow}}_0 |    \mathbf{0}_{\downarrow} , FS(N_{\uparrow}) \rangle |^2 \; ,
\label{eq:Z}
\end{equation}
with $|  \Psi^{N_{\uparrow}}_0  \rangle$ the fully interacting ground state and $|  \mathbf{0}_{\downarrow} , FS(N_{\uparrow}) \rangle$ a free spin-down atom carrying momentum $\mathbf{p}=\mathbf{0}$ in a non-interacting Fermi sea $FS$ of $N_{\uparrow}$ spin-up atoms. 
The spin-up atoms are non-interacting since $p$-wave scattering is negligible. 
The residue reflects the impurity's probability of free propagation.

The outline of the paper is as follows. In Section \ref{sec1} we introduce the model 
and the structure of the Feynman diagrammatic expansion.  
In Section \ref{sec2} we discuss  the results of the numerical calculations. Thereby, we investigate on how the results depend on the choices made with regard to the diagrammatic series, like the use of bare versus dressed propagators. Also the resummation of the diagrammatic series is discussed in depth.
The results for the quasiparticle properties, like the residue, are the subject of Section \ref{sec3}.

\section{Model and Diagrammatic structure}\label{sec1}

We consider a dilute two-component gas of ultra-cold fermionic atoms  interacting via the van der Waals-potential.  
The Hamiltonian has a kinetic and interaction term
\begin{eqnarray}
 \hat{H}   &  = &       \sum_{\mathbf{k}, \sigma = \uparrow \downarrow}  \epsilon_{\mathbf{k} \sigma}  ~ \hat{c}^{\dagger}_{\mathbf{k} \sigma} \hat{c}^{\phantom{\dagger}}_{\mathbf{k} \sigma}
\nonumber \\ 
 &   +     &   \frac{1}{\mathcal{V}} \sum_{\mathbf{k}, \mathbf{k}', \mathbf{q}}
V(\mathbf{k} - \mathbf{k}')  ~
\hat{c}^{\dagger}_{\mathbf{k}+\frac{\mathbf{q}}{2} \uparrow} \hat{c}^{\dagger}_{-\mathbf{k}+\frac{\mathbf{q}}{2} \downarrow} \hat{c}^{\phantom{\dagger}}_{-\mathbf{k}'+\frac{\mathbf{q}}{2} \downarrow} \hat{c}^{\phantom{\dagger}}_{\mathbf{k}'+\frac{\mathbf{q}}{2} \uparrow} \; .  \nonumber \\
\label{eq:ham}
\end{eqnarray}
The operators $  \hat{c}^{\dagger}_{\mathbf{k} \sigma}$   ($\hat{c}^{\phantom{\dagger}}_{\mathbf{k} \sigma}$) create (annihilate) fermions with momentum $\mathbf{k}$ and spin $\sigma$. The spin-$\sigma$ fermions have mass $m_{\sigma}$ and dispersion $\epsilon_{\mathbf{k} \sigma} = k^2 / 2 m_{\sigma}$, and $\mathcal{V}$ is the volume of the system.  We take $\hbar=1$ throughout the paper, and consider the mass-balanced case $m_{\uparrow} = m_{\downarrow} = m$.
All the theoretical considerations are for zero temperature (or $T\ll T_F$ with $T_F$ the Fermi temperature).
The diluteness of the system ensures that the range $b$ of the 
potential is much smaller than the typical inter-particle distance $1/k_F$, or $k_F b \ll 1$, with $k_F$ the Fermi momentum of the spin-up sea, 
and therefore the  details of the interaction potential become irrelevant.
Accordingly, without loss of generality, one can model the short-ranged interaction as a contact interaction, $V(\mathbf{r}) = g_0 \delta(\mathbf{r})$,
in combination with the standard ultra-violet divergence regularization procedure described below. 

The one-body and two-body propagators provide access to information about the `polaron' and `molecule' channel respectively.  
The polaron and molecule are two distinct objects belonging to different charge sectors. The one-body and two-body propagators are discussed in Sections \ref{subseca} and \ref{subsecc}.
The adopted regularization procedure for the renormalized interaction is the subject of Section  \ref{subsecb}.
The DiagMC method is introduced in Section \ref{subsecd}.

\subsection{One-body propagator}\label{subseca}

The polaron quasiparticle properties can be extracted from the impurity's Green's function defined as
\begin{equation}
G_{\downarrow}(\mathbf{k}, \tau) =  -   \theta(\tau)    \langle \Phi_0^{N_{\uparrow}} |   
 \hat{c}^{\phantom{\dagger}}_{\mathbf{k} \downarrow} (\tau)   \hat{c}^{\dagger}_{\mathbf{k} \downarrow}(0) 
    ~  | \Phi_0^{N_{\uparrow}}  \rangle   \; ,
\label{eq:Gdown}
\end{equation}
with $\hat{c}^{\phantom{\dagger}}_{\mathbf{k} \downarrow} (\tau)$ the annihilation operator in the Heisenberg picture,
 \begin{equation}
 \hat{c}^{\phantom{\dagger}}_{\mathbf{k} \downarrow} (\tau)  = e^{(\hat{H}-\mu \hat{N}_{\downarrow}-\mu_{\uparrow}\hat{N}_{\uparrow}) \tau}  \hat{c}^{\phantom{\dagger}}_{\mathbf{k} \downarrow}  e^{- (\hat{H}-\mu \hat{N}_{\downarrow}-\mu_{\uparrow}\hat{N}_{\uparrow}) \tau} \; .
 \end{equation}
The propagator $G_{\downarrow}(\mathbf{k}, \tau)$ is written in the momentum imaginary-time representation, 
 $\mu$ is a free parameter, $\hat{N}_{\sigma}$ is the number operator for spin-$\sigma$ particles, and $\mu_{\uparrow}$ is the chemical potential of the spin-up sea. 
The state 
 \begin{equation}
 | \Phi_0^{N_{\uparrow}}  \rangle  =  | \rangle_{\downarrow} | FS ({N_{\uparrow}} )\rangle \; ,
 \end{equation}
  consists of the spin-down vacuum and the    non-interacting    spin-up Fermi sea. 
Since we are dealing with an impurity spin-down atom, $G_{\downarrow}$ is only non-zero for times $\tau > 0$. The ground-state energy and $Z$-factor  can be 
extracted from the Green's function of Eq.~(\ref{eq:Gdown}). Inserting a complete set of eigenstates $|\Psi^{N_{\uparrow}}_n\rangle$ of the full Hamiltonian (\ref{eq:ham})
for one spin-down particle and $N_{\uparrow}$ spin-up particles
into Eq.~(\ref{eq:Gdown}) yields for $\mathbf{k}=\mathbf{0}$
 \begin{eqnarray}
 G_{\downarrow}(\mathbf{0}, \tau)  & = &   - \theta(\tau) \sum_n   
 | \langle \Psi^{N_{\uparrow}}_n |  \hat{c}^{\dagger}_{\mathbf{0} \downarrow} | \Phi_0^{N_{\uparrow}} \rangle  |^2 \nonumber \\
  & & \times ~ e^{- ( E_n(N_{\uparrow}) -  E_{FS}-\mu) \tau} \nonumber \\
 & \overset{\tau \to +\infty}{=} &   - Z_p  ~ e^{-(E_p-\mu) \tau} \;, 
 \label{eq:Gdownlehman}
 \end{eqnarray}
 with $E_p$ the energy of the polaron, $E_n(N_{\uparrow})$ the energy eigenvalues of the Hamiltonian (\ref{eq:ham}) and 
 $E_{FS} = 3 ~\epsilon_F N_{\uparrow} / 5$ the energy of the ideal spin-up Fermi gas, with $\epsilon_F = k_F^2/(2m)$ the Fermi energy.

The difference between the polaronic and molecular state is embedded in the factors $ | \langle \Psi^{N_{\uparrow}}_n |  \hat{c}^{\dagger}_{\mathbf{0} \downarrow} | \Phi_0^{N_{\uparrow}} \rangle  |^2$ in Eq.~(\ref{eq:Gdownlehman}).  For situations where the polaron is a well-defined quasiparticle in the ground state $|\Psi^{N_{\uparrow}}_0\rangle$, we have Eq.~(\ref{eq:Z}) for the $Z$-factor and $E_p = E_0(N_{\uparrow}) - E_{FS}$. 
 If, on the other hand, the ground state $|\Psi^{N_{\uparrow}}_0 \rangle$  is a dressed molecule the overlap $ \langle \Psi^{N_{\uparrow}}_0 |  \hat{c}^{\dagger}_{\mathbf{0} \downarrow} | \Phi_0^{N_{\uparrow}} \rangle $  is zero \cite{punk}. This is clear from the expansion of the molecular state in the number of particle-hole excitations, 
 \begin{eqnarray}
 & & |\Psi^{N_{\uparrow}}_0 \rangle   =    \bigg(     \sideset{}{'}\sum_{\mathbf{k}}  \xi_{\mathbf{k}}     \hat{c}^{\dagger}_{-\mathbf{k} \downarrow}     \hat{c}^{\dagger}_{\mathbf{k} \uparrow}  
  \nonumber \\
 & &   +       \sideset{}{'}\sum_{\mathbf{k},\mathbf{k}',\mathbf{q}}     \xi_{\mathbf{k}\mathbf{k}'\mathbf{q}}       \hat{c}^{\dagger}_{\mathbf{q}-\mathbf{k}-\mathbf{k}' \downarrow}     \hat{c}^{\dagger}_{\mathbf{k} \uparrow}    \hat{c}^{\dagger}_{\mathbf{k}' \uparrow}     \hat{c}^{\phantom{\dagger}}_{\mathbf{q} \uparrow}  
 + \ldots   \bigg)  | \Phi_0^{N_{\uparrow}-1} \rangle  \; .~~~
 \end{eqnarray}
The coefficients $\xi$ are variational parameters, and the primes indicate that the sums on $\mathbf{k}$, $\mathbf{k}'$ and $\mathbf{q}$ are restricted to $|\mathbf{k}|, |\mathbf{k}'| > k_F$ and $|\mathbf{q}|<k_F$.
Even if a molecule is formed in the ground state, the polaron can  be a well-defined excited state (in the sense of a narrow peak in the spectral function), and $Z_p$ can be non-zero.

 For vanishing interactions $V$ the impurity Green's function of Eq.~(\ref{eq:Gdown}) becomes
 \begin{equation}
 G^{0}_{\downarrow}(\mathbf{k}, \tau) = - \theta(\tau) e^{- (\epsilon_{\mathbf{k} \downarrow} -\mu) \tau} \; . 
 \end{equation}
 The one-body propagator for the spin-up sea is defined as
 \begin{equation}
 G_{\uparrow}(\mathbf{k}, \tau) =  -     \langle \Psi^{N_{\uparrow}}_0 |   
T_{\tau} \big[ \hat{c}^{\phantom{\dagger}}_{\mathbf{k} \uparrow} (\tau)   \hat{c}^{\dagger}_{\mathbf{k} \uparrow}(0) \big]
    ~  | \Psi^{N_{\uparrow}}_0 \rangle   \; ,
\label{eq:Gup}
 \end{equation}
 with $T_{\tau}$ the time-ordering operator. Without interactions, one obtains the free propagator
 \begin{equation*}
 G^{0}_{\uparrow}(\mathbf{k}, \tau) = \left \{
 \begin{array}{rl}
  -   & e^{-(\epsilon_{\mathbf{k}}-\epsilon_F)\tau}  \theta(|\mathbf{k}|-k_F)   ~~ {\rm if } ~~ \tau > 0 \; , \\
    & e^{-(\epsilon_{\mathbf{k}}-\epsilon_F)\tau}   \theta(k_F-|\mathbf{k}|) ~~{\rm if } ~~ \tau < 0 \; .
  \end{array} \right.
  \label{eq:G0up}
 \end{equation*}
 
\begin{figure}
\includegraphics[angle=-90, width=5cm] {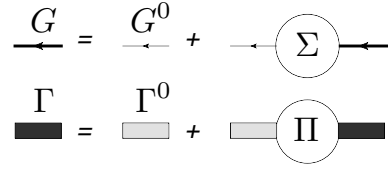}
\caption{\label{fig:dyson}  Graphical representation of the Dyson equation and the Bethe-Salpeter equation. The free (dressed) one-body impurity propagator is denoted by $G_{\downarrow}^0$ ($G_{\downarrow}$). The $\Sigma$ and $\Pi$ are the one-body and two-body self-energies, respectively. The $\Gamma$ is the fully dressed interaction, wheres $\Gamma^0$ is the partially dressed interaction obtained by summing all the bare ladders $G^0_{\downarrow} G^0_{\uparrow}$ (see Eqs. (\ref{eq:DysonGamma0tilde}) and (\ref{eq:pi0})). }
\end{figure}

 Our goal is to calculate the $G_{\downarrow}$ of Eq.~(\ref{eq:Gdown}) to extract $E_p$ by means of Eq.~(\ref{eq:Gdownlehman}).   This is achieved by summing all irreducible one-particle self-energy diagrams with the  DiagMC algorithm (which works in momentum-imaginary-time representation). 
 The irreducible self-energy $\Sigma(\mathbf{k}, \omega)$ in imaginary-frequency representation is obtained after a numerical Fourier transform, and
 inserted into Dyson's equation to give $G_{\downarrow}$, 
 \begin{equation}
 \big[ G_{\downarrow} (\mathbf{k}, \omega) \big]^{-1}= \big[ G^{0}_{\downarrow}  (\mathbf{k}, \omega)\big]^{-1} - \Sigma(\mathbf{k}, \omega) \; ,
\label{eq:Dyson}
 \end{equation}
 with $\omega$ the imaginary frequency. 
 A graphical representation of the Dyson equation is shown in the top panel of Fig. \ref{fig:dyson}.
 As was shown in Ref. \cite{polaron2}, the polaron energy $E_p$ and $Z$-factor $Z_p$ can be extracted directly from the self-energy $\Sigma(\mathbf{0},\tau)$,
 \begin{eqnarray}
 E_p & = & \int_0^{+\infty} d\tau ~ \Sigma(\mathbf{0}, \tau) ~e^{(E_p-\mu)\tau} \label{eq:ep} \; ,\\
 Z_p & = & \frac{1}{1 - \int_0^{+\infty} d\tau ~ \tau~ \Sigma(\mathbf{0}, \tau) ~e^{(E_p-\mu)\tau}} \; .
 \end{eqnarray}
 The effective mass $m_*$ of the polaron is evaluated with the estimator\cite{polaron2}
 \begin{eqnarray}
 m_*  = \frac{1/Z_p}{1/m+B_0} \; ,
 \end{eqnarray}
 with
 \begin{eqnarray}
B_0 = \int_0^{+\infty} d\tau ~ e^{(E_p-\mu)\tau}  ~ \bigg[ \frac{1}{3} \nabla^2_{\mathbf{k}}\Sigma(\mathbf{k}, \tau) |_{k=0} \bigg] \; ,
 \end{eqnarray}
 which can conveniently be estimated 
  by expanding $\Sigma(\mathbf{k},\tau)$ in Legendre polynomials. One obtains
 \begin{eqnarray}
  \frac{1}{3} \nabla^2_{\mathbf{k}}\Sigma(\mathbf{k}, \tau) |_{k=0} =    \frac{15}{2 \Delta^3} \int_0^{\Delta} dk ~\Sigma(k, \tau)   \bigg(\frac{3k^2}{\Delta^2}-1\bigg) \; ,
 \end{eqnarray} 
 and the integral can be evaluated during the MC simulation. The upper limit of integration ($\Delta$) is optimized to minimize the statistical noise while avoiding a systematic error at too large $\Delta$.
  We also used an alternative way by calculating the quasiparticle spectrum $E(\mathbf{k})$ and fitting $m_*$ via $E(\mathbf{k}) = E_p + k^2/(2m_*)$.

\subsection{Renormalized interaction} \label{subsecb}

We introduce the $s$-wave scattering length $a$ for collisions between spin-up and spin-down particles. 
 One of the advantages of working with Feynman diagrams is that one can work directly in the zero-range limit 
$k_F b \to 0$ (or, equivalently, $ \Lambda/k_F \to +\infty$ with $\Lambda$ an ultraviolet momentum cut-off) while keeping $k_F a$ constant. 
Thereby, the ultra-violet physics can be taken into account by means of a summation over all Feynman ladder diagrams.

 In momentum-imaginary-frequency representation $(\mathbf{p}, \Omega)$, 
one obtains for the partially dressed interaction
\begin{eqnarray}
\Gamma^0(\mathbf{p}, \Omega)  & =&  g_0 + g_0  ~\Pi^0(\mathbf{p}, \Omega)  ~ \Gamma^0(\mathbf{p}, \Omega ) \; ,
\label{eq:DysonGamma0tilde}
\end{eqnarray}
with $\Pi^0$ the two-particle self-energy consisting of one `bare' ladder
\begin{eqnarray}
\Pi^0(\mathbf{p}, \Omega)  =  - \frac{1}{2 \pi \mathcal{V}}  \sum_{|\mathbf{q}|<\Lambda}  \int d\omega ~G^0_{\uparrow}(\frac{\mathbf{p}}{2}+\mathbf{q}, \omega)  \nonumber \\ 
 \times ~G^0_{\downarrow}(\frac{\mathbf{p}}{2}-\mathbf{q}, \Omega - \omega) \nonumber \\
  =     \frac{1}{\mathcal{V}}  \sum_{|\mathbf{q}|<\Lambda}  ~  \frac{ \theta(|\mathbf{p}/2+\mathbf{q}|-k_F)}{i \Omega  - p^2 / (4 m ) - q^2/m + \mu + \epsilon_F}  \; ,
\label{eq:pi0}
\end{eqnarray}
where the momentum cutoff $\Lambda$ is required to keep the sum  finite.
The bare coupling constant $V(\mathbf{p}) = \int d\mathbf{r} e^{-i\mathbf{p} \cdot \mathbf{r}} V(\mathbf{r}) =  g_0$ can be eliminated in favor of the physical scattering length $a$ by using standard scattering theory
\begin{equation}
\frac{1}{g_0} = \frac{m}{4 \pi a} - \frac{1}{\mathcal{V}} \sum_{|\mathbf{k}|<\Lambda} \frac{1}{2 \epsilon_{\mathbf{k}}} \; .
\end{equation}
The
$\Gamma^0(\mathbf{p}, \Omega)$ from Eq.~(\ref{eq:DysonGamma0tilde}) can be expressed in terms of the $s$-wave scattering length $a$, by taking the limit $\Lambda \to +\infty$ and $g_0 \to 0^-$ with $a$ fixed.  In this zero-range limit, one gets
\begin{eqnarray}
[ \Gamma^0(\mathbf{p}, \Omega) ]^{-1}  = [ \tilde{\Gamma}^0(\mathbf{p}, \Omega) ]^{-1} 
 - \bar{\Pi}(\mathbf{p}, \Omega) 
\label{eq:Gamma0tilde}
\; ,
\end{eqnarray}
with 
\begin{eqnarray}
 \bar{\Pi}(\mathbf{p}, \Omega)  & = &  - \ \int \frac{d{\mathbf{q}}}{(2\pi)^3} ~ \theta(k_F-q) \nonumber \\
 & & \times \frac{1}{ i \Omega - \frac{q^2}{2m} -\frac{ (\mathbf{p}-\mathbf{q})^2}{2m}  +\mu +\epsilon_F} \; .
\label{eq:pibar}
\end{eqnarray}
Here, we have taken the thermodynamic limit ($\mathcal{V} \to +\infty$ and $N_{\uparrow}/\mathcal{V}$ fixed).
The integral in Eq. (\ref{eq:pibar}) can be evaluated analytically, 
and the dressed interaction in vacuum is given by
\begin{equation}
[ \tilde{\Gamma}^0(\mathbf{p}, \Omega) ]^{-1}  =  \frac{m}{4 \pi a} - \frac{m}{8\pi} \sqrt{p^2 - 4 m (i \Omega + \mu + \epsilon_F) }  \; , 
\end{equation}
for $\Omega\neq0$ or $\mu < -\epsilon_F$, and assuming the principal branch. For $\mu < - [\epsilon_F + 1/(ma^2)]$, the Fourier transform to imaginary time
 can be done analytically, producing
\begin{eqnarray}
& & \tilde{\Gamma}^0(\mathbf{p},\tau)  =    - \frac{4\pi}{m^{3/2} }   e^{-(\frac{p^2}{4m}-\mu-\epsilon_F)\tau} \nonumber \\  
& & \times \bigg(\frac{1}{\sqrt{\pi \tau}} + \frac{1}{\sqrt{m} a} e^{\frac{\tau}{ma^2}} 
{\rm{erfc}}\Big(-\sqrt{\frac{\tau}{m}} \frac{1}{a}\Big) \bigg) 
 \; ,
\end{eqnarray}
 with ${\rm erfc}(x)$ the complementary error function. As in Ref. \cite{polaron2}, we use $\Gamma^0(\mathbf{p},\tau)$ as a partially dressed interaction vertex in the diagrammatic series, instead of the bare interaction vertex $g_0$. This dressed vertex is calculated 
 here in imaginary time representation by performing the Fourier transform of Eq.~(\ref{eq:Gamma0tilde}) numerically.

 In a next step, the interaction vertex will be fully dressed by calculating the two-particle self-energy $\Pi$ and plugging it into the Bethe-Salpeter equation, 
 \begin{equation}
[ \Gamma(\mathbf{p}, \Omega) ]^{-1} = [ \Gamma^0(\mathbf{p}, \Omega) ]^{-1} - \Pi(\mathbf{p}, \Omega) \; . 
 \label{eq:BS}
 \end{equation}
 A graphical representation of this equation is shown in Fig.~\ref{fig:dyson}.
The self-energy $\Pi$ contains all  connected two-particle diagrams that are irreducible with respect to cutting a single $\Gamma^0$ propagator. To avoid double counting, the diagrams for $\Pi$ should not contain any ladders, since those have been summed in $\Gamma^0$ by means of the Eq.~(\ref{eq:DysonGamma0tilde}). This rule also holds when summing diagrams for the one-body self-energy $\Sigma$,  built from free propagators $G^{0}_{\sigma}$ and $\Gamma^0$.

\subsection{Two-body propagator}\label{subsecc}

Here, we consider the pair annihilation operator, 
\begin{equation}
\hat{P}_{\mathbf{k}} = \sum_{\mathbf{q}} \varphi(\mathbf{q}) ~
\hat{c}^{\phantom{\dagger}}_{\mathbf{k}-\mathbf{q} \uparrow} \hat{c}^{\phantom{\dagger}}_{\mathbf{q} \downarrow} \; ,
\label{eq:pair}
\end{equation}
with $\varphi(\mathbf{q})$ the momentum representation of the wave function $\varphi(\mathbf{r})$
for relative motion of the two fermions of opposite spin.
The two-particle propagator is defined as
\begin{equation}
 G_2(\mathbf{k}, \tau) = -   \theta(\tau)    \langle \Phi_0^{N_{\uparrow}} |   
 \hat{P}^{\phantom{\dagger}}_{\mathbf{k} } (\tau)   \hat{P}^{\dagger}_{\mathbf{k} }(0) 
    ~  | \Phi_0^{N_{\uparrow}} \rangle   \; ,
\end{equation}
where we included the fact that the impurity spin-$\downarrow$ atom  propagates forward in time.
Inserting the complete basis $|\Psi^{N_{\uparrow}+1}_n\rangle$ for ($N_{\uparrow}+1$) spin-up particles and one spin-down particle, gives 
\begin{eqnarray}
 G_2(\mathbf{0}, \tau) &  = &   -   \theta(\tau)  \sum_n   
 | \langle      \Psi^{N_{\uparrow}+1}_n |    \hat{P}^{\dagger}_{\mathbf{0} }
   ~  | \Phi_0^{N_{\uparrow}} \rangle  |^2  \nonumber \\
& &     \times ~e^{-(E_n(N_{\uparrow}+1)-E_{FS}(N_{\uparrow}) -\mu_{\uparrow}-\mu) \tau}  \nonumber \\ &
 \overset{\tau \to +\infty}{=} &   - Z_{\rm{mol}}  ~ e^{-(E_{\rm{mol}}-\mu) \tau} \;, 
\end{eqnarray}
with $E_{\rm mol}$ the molecule energy and $Z_{\rm{mol}}$ the molecule $Z$-factor. If the molecule is a well-defined quasiparticle in the ground state, we have
\begin{equation}
Z_{\rm{mol}} =  | \langle      \Psi^{N_{\uparrow}+1}_0 |    \hat{P}^{\dagger}_{\mathbf{0} } | \Phi_0^{N_{\uparrow}} \rangle  |^2 \; ,
\end{equation}
and $E_{\rm{mol}} = E_0(N_{\uparrow}+1)-E_{FS}(N_{\uparrow}) -\mu_{\uparrow}$. 
Note that the value of $Z_{\rm{mol}}$ depends on the wave function $\varphi(\mathbf{q})$. 
The functional from of this pair wave function depends on the nature of experiment used to probe the molecule.

In practice,  it is easier to calculate the molecule energy from the fully dressed interaction $\Gamma$ (see Eq.~(\ref{eq:BS})). This function is closely related to the pair correlation function, namely, 
\begin{equation}
\Gamma(\mathbf{k},\tau) = g_0 \delta(\tau) + g_0~ \mathcal{P}(\mathbf{k},\tau)~ g_0 \; ,
\label{eq:GammaP}
\end{equation}
with
\begin{equation}
\mathcal{P}(\mathbf{r},\tau)  = -      \theta(\tau)    \langle \Phi_0^{N_{\uparrow}} |          (\hat{\Psi}_{\uparrow} \hat{\Psi}_{\downarrow} ) (\mathbf{r},\tau)    ( \hat{\Psi}^{\dagger}_{\downarrow}  \hat{\Psi}^{\dagger}_{\uparrow} ) (\mathbf{0},0)       |  \Phi_0^{N_{\uparrow}}\rangle \; ,
\label{eq:pcorr}
\end{equation}
the pair correlation function.   
The field operators $ \hat{\Psi}^{\dagger}_{\sigma}(\mathbf{r}) = \sum_{\mathbf{k}} e^{-i\mathbf{k} \mathbf{r}} \hat{c}^{\dagger}_{\mathbf{k},\sigma} /\sqrt{\mathcal{V}}$ create a spin-$\sigma$ fermion at position $\mathbf{r}$. In Eq.~(\ref{eq:pcorr}) the pair of particles is created at the same position (which corresponds to $\varphi(\mathbf{q})=1$ in Eq.~(\ref{eq:pair})).
The structure of 
the fully dressed interaction $\Gamma$ and the two-particle propagator $G_2$ now implies that both structures have the same poles (see Eq.~(\ref{eq:GammaP})). Therefore, the exponential tail of the function $\Gamma(\mathbf{k}=\mathbf{0},\tau)$ can conveniently be used for estimating the molecule energy, rather than the tail of $G_2(\mathbf{0},\tau)$. This is equivalent with looking for this pole of the Bethe-Salpeter equation (\ref{eq:BS}). The molecule's energy $E_{\rm mol}$ is given by the parameter $\mu$ that satisfies the equation
\begin{equation}
[ \Gamma^0(\mathbf{p}=\mathbf{0}, \Omega=0) ]^{-1} =  \Pi(\mathbf{p}=\mathbf{0}, \Omega=0) \; ,
\end{equation}
where the left-hand-side is known analytically, and the right-hand-side is evaluated with the DiagMC algorithm in imaginary-time domain.

\subsection{Diagrammatic Monte Carlo}\label{subsecd}

\begin{figure}
\includegraphics[angle=-90, width=0.9\columnwidth] {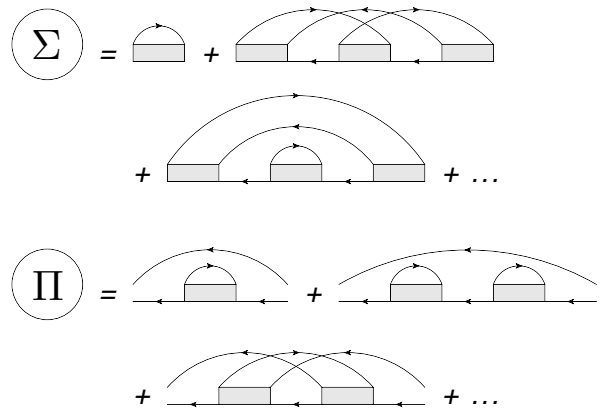}
\caption{\label{fig:series_bare}  Diagrammatic expansion for the one-body self-energy $\Sigma$ and the two-body self-energy $\Pi$. Here, the diagrams are built from the bare propagators $G^0_{\sigma}$ (thin lines), and the partially dressed interaction $\Gamma^0$ (light grey box). All diagrams have a `backbone' structure, since we have a single impurity propagating forward in time and interacting with a Fermi sea of free particles. }
\end{figure}

DiagMC evaluates the series of Feynman diagrams for the self-energy in a stochastic way.  
We deal with both the one-body and two-body self-energies.
In a first step, 
the self-energy is built from the free propagators $G^{0}_{\sigma}$ and the partially dressed interaction $\Gamma^0$ (obtained through summation of $G^{0}_{\downarrow} G^{0}_{\uparrow}$ ladders; as discussed in Section~\ref{subsecb}).   
We will refer to this series as the `bare series'. 
Fig.~\ref{fig:series_bare} shows the  one-body and two-body self-energy diagrams up to order $3$ in the bare scheme. 
The order of a diagram is  $N$ when there are $N$  dressed interactions $\Gamma$ (i.e., $N$ boxes) present in the $\Sigma$-diagram, and $N-1$ boxes in the $\Pi$-diagram.
Note that the diagrams cannot contain ladders since these have been taken into the vertex function $\Gamma^0$.
To illustrate the factorial growth with order, the number of one-body self-energy diagrams for given order $N \leq 12$  is given in the  second column of Table~\ref{tab:numbers} for the bare series. 

\begin{figure}
\includegraphics[angle=-90, width=0.9\columnwidth] {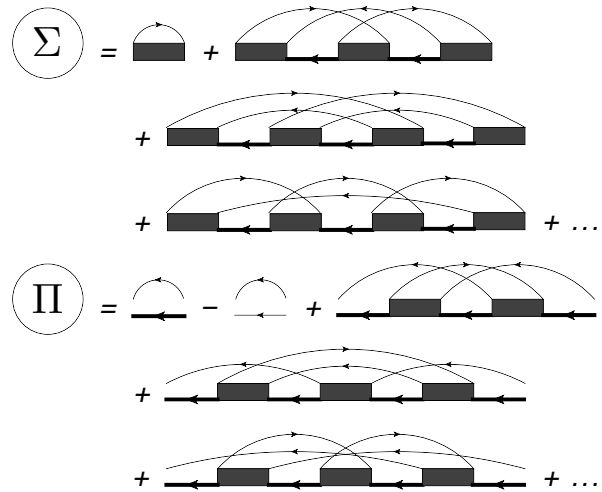}
\caption{\label{fig:series_bold}  
Skeleton diagrammatic expansion for the one-body and two-body self-energy: the impurity propagator and interaction  lines that appear in the diagram are fully dressed solutions of the Dyson equation and the Bethe-Salpeter equation (see Fig.~\ref{fig:dyson}).
}
\end{figure}

In a second step, we will use dressed propagators or `bold lines'  in the diagrams. 
Such dressed (skeleton) series are evaluated with the Bold DiagMC technique \cite{bold,polaron1}.
We consider the case with only dressed $G_{\downarrow}$ propagators while keeping $\Gamma^0$ of Eqs.~(\ref{eq:Gamma0tilde}) and (\ref{eq:pibar}) as renormalized interaction, and the case whereby both the one-body propagators and interactions 
 are dressed.
We will refer to these skeleton series as `bold $G$' and `bold $G$-$\Gamma$', respectively.
 In the latter case, the Bold DiagMC algorithm is constructed as follows:
 given approximate one-body and two-body self-energies $\Sigma$ and $\Pi$, the Dyson and Bethe-Salpeter equation are solved to deliver the one-body propagator $G_{\downarrow}$ and  the dressed interaction $\Gamma$ (see Eqs. (\ref{eq:Dyson}) and (\ref{eq:BS})). In a next step, these are used to dress the series for $\Sigma$ and $\Pi$, which are evaluated stochastically with DiagMC up to order $N_*$. This  self-consistent cycle is repeated until convergence is reached.  
 Fig.~\ref{fig:series_bold} shows the skeleton (bold $G$-$\Gamma$) series for the one- and two-body self-energies up to order 4.   Evidently, when dressing the lines in the self-energies, one has to keep track of two-particle reducibility, and systematically avoid any double counting. 
 This typically means that at any order $N$ the numbers in the second column of Table~\ref{tab:numbers} (Bare) are an upper limit of the number of diagrams in the third and fourth column.
  At $N=2$ and $N=4$, however,  the number of diagrams increases due to the fact that ladders should be included again once $G_{\downarrow}$ is bold. 
 All the diagrams of Table~\ref{tab:numbers}  are summed explicitly during  the (Bold) DiagMC simulation. 

\begin{table}
\begin{tabular}{lrrr}
\hline
$N$ &
\multicolumn{1}{r}{Bare} &
\multicolumn{1}{r}{Bold $G$} &
\multicolumn{1}{r}{Bold $G$-$\Gamma$}\\ 
\hline
1       &      1 & 1 & 1 \\
2       &       0 & 1 & 0  \\
3	& 		2				&	2					&	1	 \\
4	&		6				&	7					&	2 \\
5	&		34				&	34					&	13 \\
6	&		210				&	206					&	74 \\
7	&		1,526				&	1,476					&	544	 \\
8	&		12,558			&		12,123			&			4,458 \\
9	&		115,618			&		111,866			&			41,221 \\
10	&		1,177,170 		&		1,143,554		&			421,412 \\
11	&		13,136,102	  		&		12,816,572		&			4,722,881	 \\
12	&	~~~~159,467,022 	&		~~~~156,217,782		&		~~~~	57,553,440 \\
\hline
\end{tabular}
\caption{
Factorial increase of the number of Feynman diagrams. At fixed order $N$, the number of one-body self-energy diagrams is given for different types of series: the bare series, the skeleton series with dressed $G_{\downarrow}$ (bold $G$), and  the skeleton series with dressed $G_{\downarrow}$ and $\Gamma$ lines (bold $G$-$\Gamma$). 
}
\label{tab:numbers}
\end{table}

\section{Resummation and boldification} \label{sec2}

\begin{figure}
\includegraphics[angle=0, width=0.9\columnwidth] {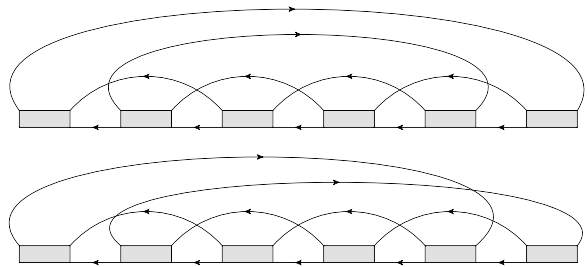}
\caption{\label{fig:cancel} 
The figure contains the two dominant  one-body self-energy diagrams  for $N=6$. Imaginary time runs from right to left. 
}
\end{figure}

\begin{figure}
\includegraphics[angle=0, width=\columnwidth] {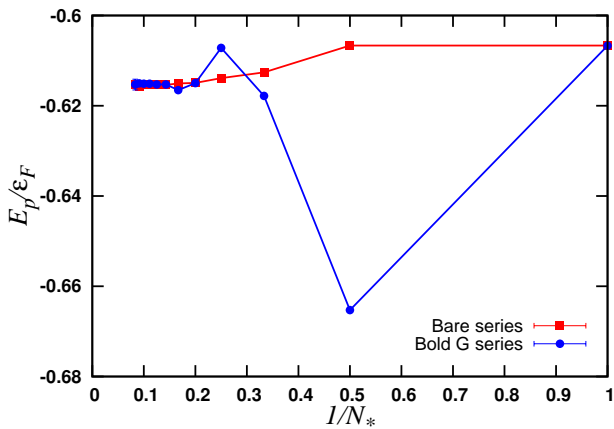}
\caption{\label{fig:Ep_ainf}  (color online) The polaron energy in units of the Fermi energy as a function of the inverse
maximum diagram order $1/N_*$ for irreducible self-energy diagrams at unitarity $1/(k_F a) = 0$.
The red squares show the polaron energy calculated via Eq.~(\ref{eq:ep}) with self-energy diagrams built from the free propagators $G^0_{\sigma}$
and the partially dressed propagator $\Gamma^0$. The blue circles show the results from the  bold-$G$ approach.
} 
\end{figure}

When considering a diagrammatic series, it is natural to ask whether  there are dominant classes of diagrams. 
Identification of  the dominant diagrams potentially allows one to make good approximations. 
To address this issue, we constructed a histogram counting how many times a certain topology is sampled. 
We consider first the bare series. %
It turns out that for the one-body self-energy, roughly half of the simulation time is spent on sampling two diagrams at each order. These two diagrams are shown in Figure \ref{fig:cancel} for diagram order six.
To understand why these two diagrams are dominant at a fixed $N$,
we use an argument 
 first made by Hugenholtz\cite{Hugenholtz}.  
For the dilute spin-up gas, momentum integration inside the Fermi sea is heavily restricted in phase space (momentum integration runs up to the Fermi momentum $k_F \sim (N_{\uparrow}/\mathcal{V})^{1/3}$). This implies the presence of a backward (or hole) spin-up
 propagator reducing the contribution of the diagram significantly, while the forward (particle) propagator enhances the contribution roughly with a factor $\int_{|\mathbf{k}|>k_F} d\mathbf{k}$. 
As a consequence, diagrams with the smallest possible number of hole propagators will be dominant. For the self-energy, we see that, at fixed order, the minimum number of hole propagators is 
two. 
Since the number of fermion loops differs by one, 
these two diagrams  have opposite sign. 
Numerically we found that the two diagrams almost cancel each other. This can be seen in Figure \ref{fig:Ep_ainf}, where we show the polaron energy $E_{p}$ as a function of the inverse diagram order cut-off $N_*$ for the interaction strength $1/(k_F a)=0$. For the `bare series', we observe a fast convergence due to cancellation of diagrams.  This magic cancellation was referred to as 
`sign blessing' \cite{polaron1}.
At infinite scattering length, such near cancellation was also observed by Combescot and Giraud \cite{Combescot}. 
 They have found that the success of the Chevy ansatz at strong coupling can be attributed to
  a nearly perfect destructive interference of the states with more than one particle-hole excitation. 
  Combescot and Giraud illustrated  that an expansion in powers of the hole wave vectors converges extremely rapidly at unitarity.  In our case, the series is 
organized differently, but at fixed order we have exactly the same type of cancellation between diagrams with the same number of hole propagators\cite{Giraudthesis}. Just like in the Combescot-Giraud argument, the cancellation is exact when the momentum-dependence of the hole propagators is neglected. Note that the dominant diagrams (see Figure \ref{fig:cancel}) 
can also be viewed as
three-body $T$-matrix diagrams closed with two hole propagators \cite{t3}. 
This class of diagrams, in which there are at most two particle-hole excitations, has been considered previously for the polaron problem \cite{mol3,Combescot}. It was shown that they exactly reproduce the Skorniakov and Ter-Martirosian equation \cite{skor} in the BEC limit. In this strong-coupling limit, the dominant process is scattering between a dimer and a spin-up fermion, which is diagrammatically  represented by the three-body T-matrix diagrams. Away from this limit, the considered class of diagrams turns out to give a quantitatively good correction to the lowest order result. We find that this is due to their dominance, even away from the BEC limit. 

\begin{figure}
\includegraphics[angle=0, width=\columnwidth] {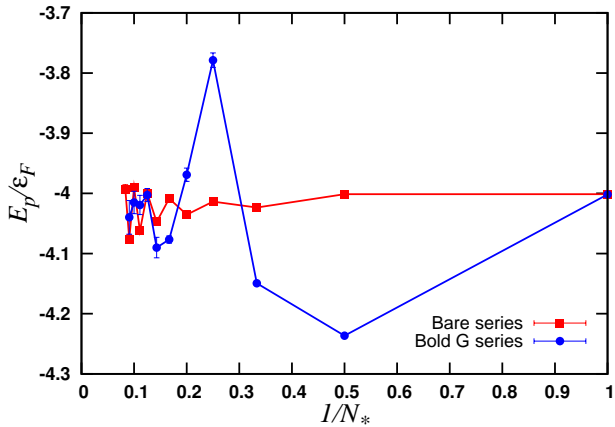}
\caption{\label{fig:E_N_a_infty3}  (color online)  Same plot as Figure \ref{fig:Ep_ainf}, but now considering the interaction strength parameter $1/(k_F a) = 1.333$.  In the bare series, small oscillations unable us to extrapolate to infinite diagram order.
} 
\end{figure}

When going towards the BEC side ($1/a>0$), the cancellation between dominant diagrams of the type shown in Figure \ref{fig:cancel} is no longer perfect. 
Figure \ref{fig:E_N_a_infty3} shows the polaron energy as a function of $1/N_*$ for $1/(k_F a) = 1.333$.  
For the bare series, the oscillations prevent one from extracting $E_p$ for $1/N_*\to 0$.

To cure the bad convergence of the bare series for $1/(k_F a)>0$ one can include more diagrams by dressing the propagators. 
We start by dressing the spin-down propagator lines, while keeping the partially dressed $\Gamma^0$. Diagrams reducible with respect to cutting \emph{two} spin-down lines should no longer be sampled, since they are included implicitly. 
For $1/(k_F a) =0$ the self-energy $\Sigma(\mathbf{k},\tau)$ converges in this `bold $G$ scheme' for $N_* \geq 7$. Extrapolation to infinite $N_*$  gives the exact  $\Sigma$ and $G_{\downarrow}$.
Figure \ref{fig:Ep_ainf} includes the polaron energy as function of diagram order cut-off when the one-body self-energy is built with the exact $G_{\downarrow}$.
The bare and bold series converge to the same energy. Remarkably, the dressed scheme gives worse results at low $N_*$. This indicates that approximations based on a few low order diagrams are completely uncontrolled, and including more diagrams by dressing the lines does not necessarily improve the quality of the results.

\begin{figure}
\includegraphics[angle=0, width=\columnwidth] {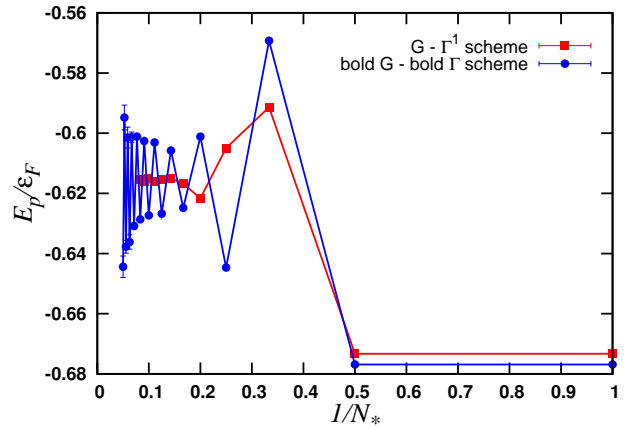}
\caption{\label{fig:E_N_a_infty2}  (color online) The polaron energy in units of the Fermi energy as a function of the inverse
maximum diagram order $1/N_*$ for irreducible self-energy diagrams at unitarity $1/(k_F a) = 0$. The blue circles show results from a fully Bold DiagMC simulation: the self-energy diagrams are built from fully dressed one-particle propagators $G_{\downarrow}$ and two-particle propagators $\Gamma$ up to self-energy diagram order $N_*$.
The red squares show the polaron energy calculated with diagrams built with the exact $G_{\downarrow}$ and a partially dressed interaction $\Gamma^1$ containing the sum of all $G_{\downarrow} G^0_{\uparrow}$ ladders. }
\end{figure}

For $1/(k_F a) = 1.333$, we see that dressing the impurity lines helps to get rid of the residual oscillations in the bare scheme (see Figure \ref{fig:E_N_a_infty3}). One might expect that dressing even more, by using a fully dressed $\Gamma$ instead of $\Gamma^0$, might lead to even better convergence. Figure \ref{fig:E_N_a_infty2} shows however that, even for $1/(k_Fa)=0$, the fully bold series (bold $G$-$\Gamma$ scheme) does not seem to converge ($N_*$ is the diagram cut-off for both $\Sigma$ and $\Pi$, and a Bold DiagMC simulation is done for each $N_*$), in  contrast to the results of Ref.\cite{polaron2}.
The data for the fully bold 
simulation of Ref.\cite{polaron2}   was obtained by using the exact $G$ and $\Gamma$ (i.e. extrapolated to the $N_* \to \infty$ limit with resummation
factors). They were not obtained with a self-consistent simulation, which explains the difference. 
Moreover, data is not shown above $N_*=7$, where oscillations do occur.
 In order to understand why the series no longer converges, we  introduce an intermediate scheme (which we call bold $G$-$\Gamma^1$): the self-energy is built from the fully converged $G_{\downarrow}$ and a partially dressed interaction $\Gamma^1$, built from summing the ladders $G_{\downarrow} G^{0}_{\uparrow}$. 
The result is also shown in Figure \ref{fig:E_N_a_infty2},  and we again observe convergence to the same answer as in Figure \ref{fig:Ep_ainf}. The key difference between both schemes is that in the bold $G$-$\Gamma^1$ scheme, both dominant diagrams shown in Fig. \ref{fig:cancel} still explicitly  contribute to the self-energy, whereas in the fully bold scheme the upper dominant diagram becomes reducible and is taken into account self-consistently. This means that the balance of cancellation between diagrams is broken, and a single dominant diagram keeps contributing at each order. So, it turns out that dressing the diagrams as much as possible is not always a good idea. In this respect, our findings disagree with Ref.\cite{polaron2}.

A second method to cure the bad convergence of the bare series on the BEC side, is to employ series resummation techniques. We will use the Abelian resummation techniques\cite{Hardy} which have been used when calculating the equation of state of  the unitary gas with Bold DiagMC\cite{vanhoucke12}. This resummation technique works as follows. One starts from a series $f(x) = \sum_n d_n x^n$ that has a \emph{finite} radius of convergence $R>0$. The idea is to sum the series at some point $x_0$ outside of the radius $R$ by analytically continuing the function $f$. This provides a good procedure for summing the divergent series in the sense that it respects basic operations (sum, multiplication and derivative) and that it preserves distinctness\cite{Hardy}. It is well-known that with analytic continuation, one can encounter problems with the existence and/or uniqueness of the solution\cite{math}. However, one can formally define a domain called the `Mittag-Leffler star' where the function can be analytically continued along straight lines $[0,x_0]$. Note that this star will always contain the disk of convergence. It can be shown\cite{Hardy} that for each point $x_0$ of the Mittag-Leffler star, the limit
\begin{equation}
\lim_{\epsilon \to 0^+} ~ \sum_n d_n x_0^n  e^{-\epsilon \lambda_n} \; ,
\end{equation} 
with $\lambda_n = n ~{\rm log}(n)$ for $n>0$ and $\lambda_0=0$, exists and is equal to the analytic continuation of $f$ to the point $x_0$. Note that within the disk of convergence the procedure works equally well, and can improve the rate of convergence. We apply the Abelian resummation technique to the expansion of  the self-energy $\Sigma$ and $\Pi$. As the analytic structure of $\Sigma$ and $\Pi$ is unknown, it is currently impossible to determine whether there is   a finite radius of convergence and whether we are in the Mittag-Leffler star. 
In practice, we apply different resummation techniques (i.e., different functions $\lambda_n$), and test the uniqueness of the result.

\begin{figure}
\includegraphics[angle=0, width=\columnwidth] {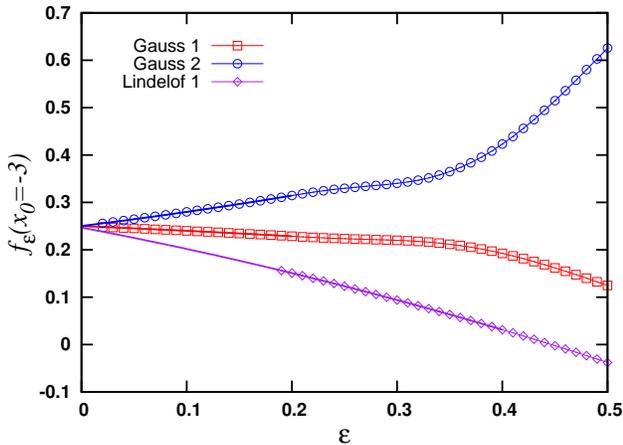}
\caption{\label{fig:resumgeom}  (color online) 
Illustration of the Abelian resummation technique for the geometric series. We evaluate $f_{\epsilon}(x_0) =  \sum_n x_0^n  e^{-\epsilon \lambda_n}$ for $x_0=-3$. and various choices of $\lambda_n$. The value of the analytically continued function $1/(1-x_0)$  is retrieved for $\epsilon\to 0^+$. 
}
\end{figure}

\begin{figure}
\includegraphics[angle=0, width=\columnwidth] {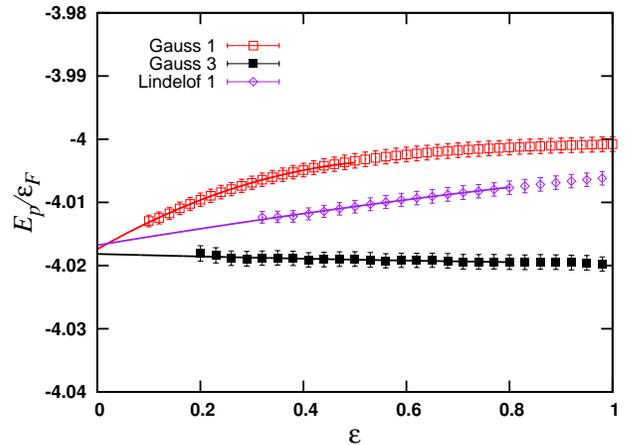}
\caption{\label{fig:resumpol}  (color online) Abelian resummation of the bare series of one-body self-energy diagrams at $1/(k_Fa)=1.333$. 
The polaron energy $E_{p}/\epsilon_F$ is extracted in the limit $\epsilon=0^+$ for different choices of $\lambda_n$.
}
\end{figure}

We  use following $\lambda_n$: (i) Lindel\"of 1: $\lambda_n  =    n ~{\rm log}(n)$ for $n>0$ and $\lambda_0=0$;  (ii) Lindel\"of 2: $\lambda_n  =    (n-1) ~{\rm log}(n-1)$ for $n>1$ and $\lambda_0=\lambda_1=0$; (iii) Gauss 1: $\lambda_n  =   n^2$  for  $n\geq0$; (iv) Gauss 2: $\lambda_n  =   (n-1)^2$  for  $n\geq1$ and  $\lambda_0=0$;   (iv) Gauss 3: $\lambda_n  =   (n-2)^2$  for  $n\geq2$ and  $\lambda_0=\lambda_1=0$.
 Before applying these resummation techniques to our diagrammatic series, we illustrate its power with an example for the geometric series. Figure \ref{fig:resumgeom} shows the sums  $f_{\epsilon}(x_0=-3) =  \sum_n x_0^n  e^{-\epsilon \lambda_n}$ for a few choices of $\lambda_n$. At small $\epsilon$, the computation of  $f_{\epsilon}$ is no longer feasible due to finite computer precision. By extrapolating to $\epsilon=0$, we indeed find $1/(1-x_0)$ with high precision. The Lindel\"of curve gives a slighly less accurate extrapolation because it suppresses high order contributions in a much smoother fashion than the Gaussian resummation. When applying these techniques to our diagrammatic series, it is the growth of the statistical error bars (due to factorial complexity) that prevents us from going to very small values of $\epsilon$.  Figure \ref{fig:resumpol} shows the polaron energy calculated with the resummed self-energy as a function of the control parameter $\epsilon$ for $1/(k_Fa)=1.333$. The polaron energy $E_p$ can be extracted with high accuracy. The major source of error bar stems from the uncertainty in the extrapolation.

\begin{figure}
\includegraphics[angle=0, width=0.9\columnwidth] {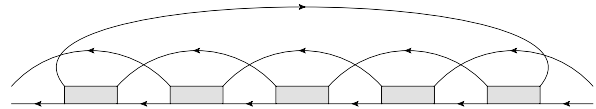}
\caption{\label{fig:diagrampol}  At fixed order, there is one dominant diagram for the two-body self-energy. Here, we draw this diagram at order six.
 }
\end{figure}

\begin{figure}
\includegraphics[angle=0, width=\columnwidth] {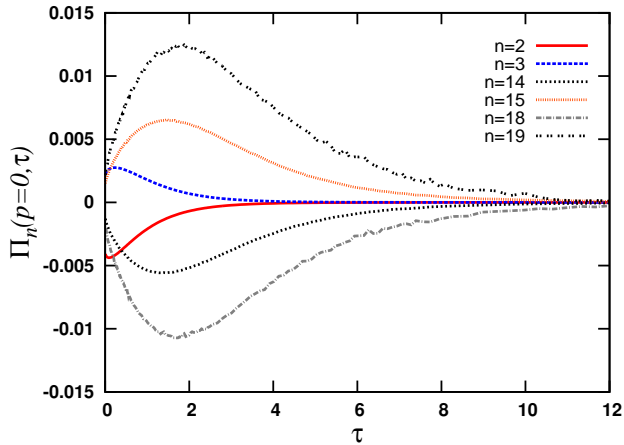}
\caption{\label{fig:Pi}  (color online) The two-particle self-energy $\Pi$ at external momentum zero as a function of imaginary time  for $k_F a = 1$. 
 The  $\Pi_n$ are the contribution of all $n$-th order diagrams, and here shown for various $n$. As $n$ increases,   $\Pi_n$  keeps on growing.  
We work in units $k_F=1$, $m=1$, $\hbar=1$ and  $\mu/\epsilon_F=-3.2$.  The noise in the curves indicates the magnitude of the statistical error. 
}
\end{figure}

\begin{figure}
\includegraphics[angle=0, width=\columnwidth] {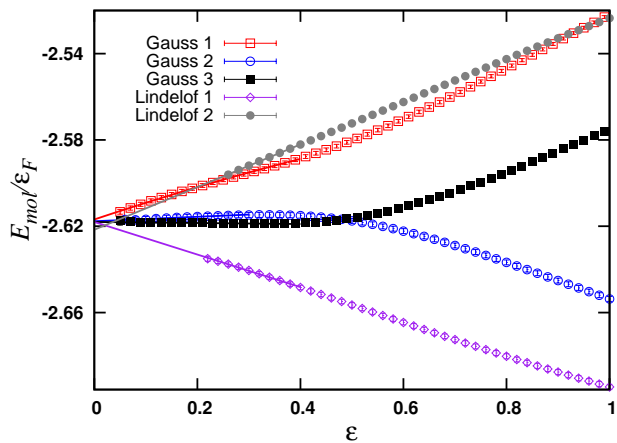}
\caption{\label{fig:resummol}  (color online) Abelian resummation of the bare series of two-particle self-energy diagrams at $k_Fa=1$. The molecule energy $E_{\rm mol}/\epsilon_F$ is extracted in the limit $\epsilon=0^+$ for different choices of $\lambda_n$.
}
\end{figure}

Histogramming the different topologies of the two-body self-energies $\Pi$ revealed a dominant diagram at each order. This diagram is shown in Figure \ref{fig:diagrampol}. Again  it shows a three-body $T$-matrix structure that is closed with a single spin-up hole propagator. Upon increasing the diagram order up to $20$, we observe a steady growth in the contribution of this diagram. This is illustrated in Figure \ref{fig:Pi}, where we plot the $n$-th order contribution $\Pi_n$ to the two-body self-energy as a function of imaginary time $\tau$ for external momentum zero.  Figure \ref{fig:resummol} illustrates that we can nonetheless get accurate values for the molecule energy $E_{\rm mol}$ by using different Abelian resummation techniques and extrapolating to $\epsilon=0^+$. Again, the Gaussian resummation methods allow one to reach very small values of $\epsilon$. 
The quoted error bars are rather conservative as we include the extrapolated results obtained with all choices for $\lambda_n$.

We also tested the resummability of the fully bold series (bold $G$ - bold $\Gamma$ scheme), since this was used in the Bold DiagMC method for determining the equation of state of the unitary gas\cite{vanhoucke12}.  When applying the Abelian resummation techniques and extrapolate to $\epsilon=0^+$ at $1/(k_F a)=0$, the correct polaron energy is retrieved. This constitutes an independent check for the resummation of the skeleton series.

\section{Quasiparticle properties} \label{sec3}

\begin{figure}
\includegraphics[angle=0, width=\columnwidth] {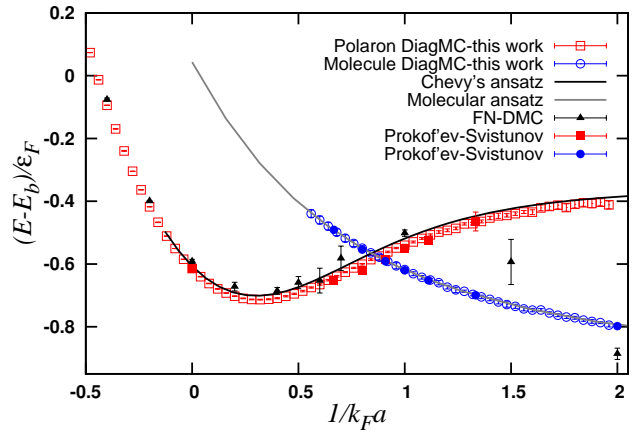}
\caption{\label{fig:ener}  (color online) 
The extracted polaron and molecule energy as a function of the interaction strength $1/(k_Fa)$. The energies are expressed as $(E-E_b)/\epsilon_F$, with $E_b=-1/(m a^2)$ the molecule energy in vacuum. The FN-DMC results are from Ref.\cite{pilati}, the variational results for the polaron from Ref.\cite{chevy} and for the molecule from Ref.\cite{punk}. DiagMC results by Prokof'ev and Svistunov\cite{polaron1} are also shown.
}
\end{figure}

As an independent cross-check of Ref. \cite{polaron1}, which uses alternate ways of resumming the diagrammatic series, we  calculate the ground-state energies of the polaron and molecule. Figure \ref{fig:ener} shows these energies shifted by the vacuum molecule energy $E_b = -1/(ma^2)$ in units of the Fermi energy $\epsilon_F$.  A selection of the polaron and molecule energies is also given in Table~\ref{tab:data}.
We find the transition point  at $(k_F a)_c = 1.15(3)$, in agreement with   Ref. \cite{polaron1}.
Close to the transition point, we find polaron energies that differ about 1\% with the polaron energies of Ref.\cite{polaron1}, which, we believe, is due to a small systematic error in the lowest order diagram in Ref.\cite{polaron1}.
The variational energies obtained from a wave-function ansatz for the polaron\cite{chevy} and the molecule\cite{punk} are very close to the Monte Carlo results.  
Note that Chevy's variational ansatz for the polaron state is completely equivalent with the non-self-consistent
$T$-matrix approximation~\cite{mol3} which is exactly our bare series at $N_*=1$.
Fixed node-difussion Monte Carlo (FN-DMC) results are also in good agreement with the DiagMC data. For $1/(k_Fa)=2$ it seems that systematic errors on the FN-DMC results were underestimated, since
FN-DMC should in principle give an upper bound to the true ground-state energy. 

\begin{figure}
\includegraphics[angle=0, width=\columnwidth] {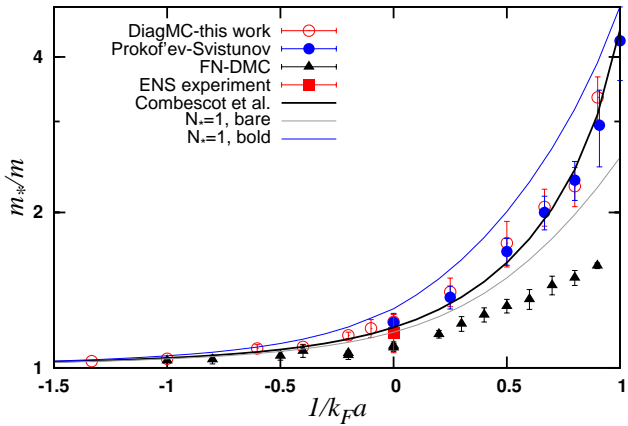}
\caption{\label{fig:mstar}  (color online) The effective mass $m_*$ of the polaron in units of the bare mass $m$ as a  function of the interaction parameter $1/(k_F a)$. Our DiagMC results (open circles) are shown together with DiagMC results by Prokof'ev and Svistunov\cite{polaron1} (filled blue circles), FN-DMC results\cite{pilati,pilatiprivate} (black triangles), ENS experiment\cite{nascim} (red square) and a variational calculation up to two particle-hole excitations \cite{mol3} (solid black line).  We also show $m_*$ calculated from the lowest order self-energy diagram (i.e., $N_*=1$) for the bare series (solid grey line) and for the fully bold $G$-$\Gamma$ series (solid blue line), which are equivalent to the non-self-consistent and the self-consistent $T$-matrix approximation, respectively.
} 
\end{figure}

Figure \ref{fig:mstar} shows the effective mass of the polaron as calculated with DiagMC. We compare with the ENS experiment \cite{nascim} at unitarity, DiagMC calculations by Prokof'ev-Svistunov\cite{polaron1}, FN-DMC\cite{pilati,pilatiprivate}, a variational calculation up to two particle-hole excitations\cite{mol3}, and the first order ($N_*=1$) result in the bare scheme and the fully bold $G$-$\Gamma$ scheme. 
The experimental effective mass, which  is in perfect agreement with DiagMC \cite{polaron1},  was extracted from the low frequency breathing modes, and in particular the Fermi polaron breathing mode. 
The lowest order bare calculation, also known as $T$-matrix approximation, is equivalent to the Chevy ansatz, while the lowest order bold calculation corresponds to the self-consistent $T$-matrix approximation.
These results show that including only single particle-hole pair excitions does not lead to accurate results for the effective mass, while the variational calculation based on  diagrams taking into account at most two  particle-hole pairs excitations agrees with the DiagMC results\cite{mol3}.

\begin{figure}
\includegraphics[angle=0, width=\columnwidth] {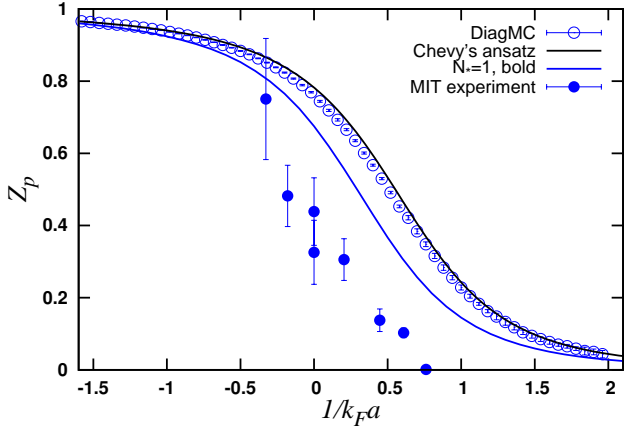}
\caption{\label{fig:zp}  (color online) The polaron quasiparticle residue $Z_p$ as a function of the interaction parameter $1/(k_F a)$. DiagMC results (open circles) are compared with variational ansatz\cite{punk} (black solid line), the fully bold $G$-$\Gamma$ series at $N_* = 1$ or self-consistent $T$-matrix approximation (solid blue line) and the MIT experiment\cite{schirotzek} (blue circles).}
\end{figure}

 Experimental and theoretical quasiparticle residues are shown in Fig.~\ref{fig:zp}. 
To create and probe polarons, the MIT experiment\cite{schirotzek} starts from a cloud of $^6$Li atoms with most atoms occupying the lowest hyperfine state $|1\rangle$ (spin-up), and about $2\%$ of the atoms occyping the hyperfine state $|3\rangle$ (spin-down) in the degenerate regime $T \approx 0.14~T_F$ with $T_F$ the Fermi temperature. 
A broad Feshbach resonance is used to enhance the scattering between atoms in states $|1\rangle$ and $|3\rangle$.
Radio-frequency (rf) spectra of the spin-up and spin-down components are measured. The atoms are transferred to a third empty state with very weak final-state interactions. Therefore, the measured transition rate $I$  can be connected with the impurity's spectral function  $\rho_{\downarrow}$ in linear response theory\cite{Ohashi05,Massignan08}
\begin{equation}
I(\omega_L) \propto \sum_{\mathbf{k}} n_F( \epsilon_{\mathbf{k}} - \mu - \omega_L) ~ \rho_{\downarrow}(\mathbf{k},   \epsilon_{\mathbf{k}} - \mu - \omega_L) \; ,
\end{equation}
with $\omega_L$ the frequency of the rf photons and $n_F(x) = 1/(1+e^{\beta x})$ the Fermi distribution. Note that the spectral function  depends on the temperature. Density inhomogeneities are taken care of through tomographic reconstruction \cite{schirotzek}.
 At sufficiently weak attractions, the Fermi polaron is  observed as a narrow peak in the impurity spectrum that is not matched by the broad environment spectrum. 
 The peak position gives the polaron energy $E_p$, and was found to be in perfect agreement with the DiagMC results of Ref. \cite{polaron1}.
 The polaron $Z$-factor was measured by determining the ratio of  the area under the impurity peak that is not matched by the environment, and the total area under the impurity's spectrum. The experimental $Z$-factor from Ref. \cite{schirotzek} is shown  in Figure \ref{fig:zp}, together with the $Z$-factor calculated from Chevy's ansatz\cite{punk,christian}, the fully self-consistent result in lowest order ($N_*=1$) and our DiagMC simulation.  DiagMC data  for the $Z$-factor is also given in Table~\ref{tab:data}. 

  The results obtained via DiagMC simulation agree extremely well with Chevy's variational ansatz. This is 
very surprising  in the strongly interacting regime where $Z_p$ is significantly smaller than one. 
Here, one would expect multiple particle-hole excitations to be important since the overlap with the non-interacting wave-function is small. Remarkably, 
including just single particle-hole excitations on top of the Fermi sea produces almost the exact $Z_p$. When the lowest order diagram is calculated in a fully self-consistent way, however, the agreement with DiagMC is less good. This hints at the fact that the almost perfect agreement with Chevy's ansatz (i.e., the lowest order bare result) is rather accidental.

 The $Z$-factors computed with Chevy's ansatz and DiagMC both exceed the measured ones. 
 It was suggested in Ref.\cite{punk} that the disagreement between the experiment and the Chevy ansatz is an artefact
of Chevy's expansion being restricted to one particle-one hole excitations. 
As the DiagMC technique includes multiple particle-multiple hole excitations and agrees very well with Chevy's ansatz, we see that this is not the case. 
However, since the measured $Z_p$ might only give a lower bound\cite{schirotzek}, theory and experiment might not be in disagreement.

The measured polaron $Z$-factor vanishes beyond a critical interaction strength. 
Ignoring issues related to metastability, once the two-body bound state becomes energetically favorable, all polarons disappear and the polaron peak vanishes. In the experiment $T/T_F=0.14(3)$, and finite-temperature effects are thus expected to become important. Indeed, close to $(k_Fa)_c=1.15(3)$ the energy difference between the molecule and polaron state is of the order of $0.1~T_F$ (Fig.~\ref{fig:ener}). Therefore, one expects that $T=0$ calculations  underestimate the critical $1/(k_Fa)$ measured at $T\approx0.1~T_F$. Similarly, the measured $1/(k_Fa)_c$ can be interpreted as an upper bound for the $T=0$ situation. 
On the other hand, due to depletion of the experimental spectrum, the measured $Z$ might only give a lower bound, which means that the experimentally determined critical  $1/(k_Fa)$ might be underestimated. These uncertainties might explain why the  critical $1/(k_Fa)$ in the experiment is lower than the value obtained with DiagMC for a single impurity. Fixed-node Monte-Carlo simulations for a finite density of impurities, on the other hand, predict phase separation before the systems even reaches the polaron-to-molecule transition\cite{pilati}, and the vanishing $Z$-factor might be a manifestation of this phase separation.

\begin{table}
\begin{tabular}{lrrr}
\hline
\multicolumn{1}{l}{$1/(k_Fa)$} &
\multicolumn{1}{r}{$E_p/E_F$} &
\multicolumn{1}{r}{$E_{\rm mol}/E_F$} &
\multicolumn{1}{r}{$Z_p$}\\ 
\hline
-1.8 &        -0.1793(1)  & &  0.9727(4)       \\
-1.6 &        -0.1961(1)& &    0.9665(5)       \\
-1.4 &        -0.2159(2)  & &  0.9590(3)        \\
-1.2 &        -0.2393(2)  & &  0.9502(3)       \\
-1.0       &  -0.2687(2)    &  & 0.9376(4)      \\
-0.8       &  -0.3052(2)   &   & 0.9209(5)       \\
-0.6	& 	-0.3526(2)					&					&	0.8978(8)     	  \\
-0.4	&	-0.4141(2)					&					&	0.8670(10)      \\
-0.2	&	-0.4976(2)					&					&	  0.8237(15)       \\
0.0	&	-0.615(1)				&							&	0.7586(27)        \\
0.2	&	-0.782(1)					&					&		 0.6720(42)       \\
0.4	&	-1.028(2)			&							&		0.5672(28)      	\\
0.6	&	-1.385(2)			&	-1.180(13) 				&		0.4410(32)      	  \\
0.8	&	-1.880(2)	 		&	-1.830(8)				&			 0.3258(58)       \\
1.0	&	-2.540(3)		  		&		-2.618(6)			&		 0.2283(70)       		  \\
1.2	&	-3.372(4) 	&		-3.554(6) 						&		 0.1559(69)       \\
1.4 &         -4.373(5)  & -4.633(5) 								&   0.1102(68)       \\
1.6 &         -5.554(8)    & -5.867(6)								 & 0.0771(58)      \\
1.8 &~~~~ -6.889(12) & ~~~~ -7.251(5) 									&~~~~ 0.0578(35)       \\
\hline
\end{tabular}
\caption{
 Selection of DiagMC data for the polaron energy $E_p$, molecule energy $E_{\rm mol}$ and polaron residue $Z_p$ for several values of the interaction strength parameter $1/(k_Fa)$.
}
\label{tab:data}
\end{table}

\section{Conclusions}

We have considered the Fermi-polaron system in three dimensions, in which a single spin-down impurity is strongly coupled to a non-interacting Fermi sea of spin-up particles. 
Although this system contains strongly interacting fermions, it can be solved  with the Diagrammatic Monte-Carlo method.
This method is based on the stochastic evaluation of a series of Feynman diagrams. 
To extract ground-state properties, one has to overcome a factorial complexity due to the increase of the number of diagrams.  
Nonetheless, extrapolation to infinite diagram order becomes possible when the diagrams cancel each other better than the factorial increase in number.
At interaction strength $1/(k_Fa)=0$, we find such perfect cancellation (within our statistical errors). 
When considering the series built on bare propagators on the BEC side, however, oscillations with diagram order remain and prevent a controlled extrapolation  to the infinite diagram order. 
We have followed two strategies around this problem: the first is to consider skeleton series (built on dressed propagators), and 
the second the use of resummation techniques. 
Though dressed series can be evaluated to higher orders, we have found that in some cases dressing can destroy a favorable cancellation of diagrams. 
For all interaction strengths we found that the (skeleton) series of the one-body and two-body self-energy is resummable by means of Abelian resummation.  
Bare series, skeleton series and resummed series give robust answers in their respective region of applicability (i.e. where the infinite diagram order extrapolation is controlled). 

We have identified classes of dominant diagrams for the one-body and two-body self-energy in the crossover region of strong interaction. The dominant diagrams turn out to be 
 the leading processes of the strong-coupling limit: scattering between a dimer and a spin-up fermion, which is diagrammatically  represented by the three-body T-matrix diagrams. Including just these dominant diagrams 
 gives a quantitatively good correction to the lowest order result, even away from the strong-coupling limit.

We have shown that not only the polaron and molecule energies agree very well with a variational ansatz from weak to strong attraction, but also the polaron residue or $Z$-factor. 
Though this agreement must be due to strong cancellation of diagrams, we only observed convergence for the bare series at $1/(k_Fa)=0$.
 A full explanation for the success of the variational ansatz 
 is still missing, and it is therefore unclear in which cases the ansatz is appropriate. 

This work is supported by the Fund for Scientific research - Flanders.
The authors would like to thank 
C. Lobo,
N. Prokof'ev, 
B. Svistunov, 
F. Werner
and 
M. Zwierlein
for the helpful discussions and suggestions.
We thank R. Combescot, S. Pilati, N. Prokof'ev, M. Punk, B. Svistunov and M. Zwierlein for sending us their data.


\begin{thebibliography}{99}
\bibitem{landau} L.D. Landau, Phys. Z. Sowjetunion {\bf 3}, 664 (1933).
\bibitem{LLStatMech2}  E.M. Lifshitz and L.P. Pitaevskii,
\textit{Statistical Mechanics}, Part 2 (Pergamon Press, New York,
1980).
\bibitem{polaron1} N.V. Prokof'ev and B.V. Svistunov, Phys. Rev. B {\bf 77}, 020408 (2008).  
\bibitem{schirotzek} A. Schirotzek, C.H. Wu, A. Sommer and M.W. Zwierlein, Phys. Rev. Lett. {\bf 102}, 230402 (2009).
\bibitem{nascim} S. Nascimb{\` e}ne, N. Navon, K.J. Jiang, L. Tarruell, M. Teichmann, J. McKeever, F. Chevy and C. Salomon,  Phys. Rev. Lett. {\bf 103}, 170402 (2009).
\bibitem{pilati} S. Pilati and S. Giorgini, Phys. Rev. Lett. {\bf 100}, 030401 (2008). 
\bibitem{pilatiprivate} S. Pilati, private communication.
\bibitem{polaron2} N.V. Prokof'ev and B.V. Svistunov, Phys. Rev. B {\bf 77}, 125101 (2008).  
\bibitem{chevy} F. Chevy, Phys. Rev. A {\bf 74}, 063628 (2006).
\bibitem{mora} C. Mora and F. Chevy, Phys. Rev. A {\bf 80}, 033607 (2009).
\bibitem{punk} M. Punk, P.T. Dumitrescu and W. Zwerger, Phys. Rev. A {\bf 80}, 053605 (2009).
\bibitem{mol3} R. Combescot, S. Giraud and X. Leyronas, EPL {\bf 88}, 60007 (2009).
\bibitem{bold} N.V. Prokof'ev and B.V. Svistunov,  Phys. Rev. Lett. {\bf 99}, 250201 (2007).   
 \bibitem{Hugenholtz} N.M. Hugenholtz, Physica {\bf 23}, 533 (1957).
\bibitem{Combescot} R. Combescot and S. Giraud, Phys. Rev. Lett. {\bf 101}, 050404 (2008).
\bibitem{Giraudthesis} S. Giraud, Ph.D. thesis, Universit\'e Paris VI, 2010, http://tel.archives-ouvertes.fr/tel-00492339 
\bibitem{t3} X. Leyronas and R. Combescot Phys. Rev. Lett. {\bf 99}, 170402, (2007).
\bibitem{skor} G.V. Skorniakov and K.A. Ter-Martirosian, Zh. Eksp. Teor. Fiz.
{\bf 31}, 775 (1956) [Sov. Phys. JETP {\bf 4}, 648 (1957)].
\bibitem{Hardy} G.H. Hardy, \textit{Divergent Series} (Oxford University Press, New York, 1956).
\bibitem{vanhoucke12} K. Van Houcke, F. Werner, E. Kozik, N. Prokofev, B. Svistunov, M.J.H. Ku, A. T. Sommer, L.W. Cheuk, A. Schirotzek and  M.W. Zwierlein, Nature Phys. {\bf 8}, 366 (2012). 
\bibitem{math} P. Dennery and A. Krzywicki, \textit{Mathematics for Physicists} (Dover, New York, 1996).
\bibitem{Ohashi05}  Y. Ohashi and A. Griffin,  Phys. Rev. A {\bf 72}, 013601 (2005).
\bibitem{Massignan08} P. Massignan, G.M. Bruun and H.T.C. Stoof, Phys. Rev. A  {\bf 77}, 031601(R) (2008).
\bibitem{christian} C. Trefzger and Y. Castin,  arXiv:1210.8179.
\end{thebibliography}
\end{document}